\documentclass[%
aps,
 pra,%
 amsmath,amssymb,
 reprint,%
 superscriptaddress,
]{revtex4-1}

\usepackage{dcolumn}

\usepackage{graphicx}
\usepackage{lineno}
\usepackage{bm}%
\usepackage[pdftex,colorlinks=true,bookmarks=false,citecolor=blue,urlcolor=blue]{hyperref}
\usepackage{upgreek}

\usepackage{titlesec} 

\begin{document}


\title{\vspace{-15mm}\fontsize{19pt}{10pt}\selectfont\textbf{Highly-tunable efficient second-harmonic generation in a lithium niobate nanophotonic waveguide}} 

\author{Rui Luo}
\affiliation{Institute of Optics, University of Rochester, Rochester, NY 14627}

\author{Yang He}
\affiliation{Department of Electrical and Computer Engineering, University of Rochester, Rochester, NY 14627}

\author{Hanxiao Liang}
\affiliation{Department of Electrical and Computer Engineering, University of Rochester, Rochester, NY 14627}

\author{Mingxiao Li}
\affiliation{Department of Electrical and Computer Engineering, University of Rochester, Rochester, NY 14627}

\author{Qiang Lin}
\email[Electronic mail: ]{qiang.lin@rochester.edu}
\affiliation{Institute of Optics, University of Rochester, Rochester, NY 14627}
\affiliation{Department of Electrical and Computer Engineering, University of Rochester, Rochester, NY 14627}

\date{\today}



\begin{abstract}
Highly-tunable coherent light generation is crucial for many important photonic applications. Second-harmonic generation (SHG) is a dominant approach for this purpose, which, however, exhibits a trade-off between the conversion efficiency and the wavelength tunability in a conventional nonlinear platform. Recent development of the integrated lithium niobate (LN) technology makes it possible to achieve a large wavelength tuning while maintaining a high conversion efficiency. Here we report on-chip SHG that simultaneously achieves a large tunability and a high conversion efficiency inside a single device. We utilize the unique strong thermo-optic birefringence of LN to achieve flexible temperature tuning of type-I inter-modal phase matching. We experimentally demonstrate spectral tuning with a tuning slope of 0.84 nm/K for a telecom-band pump, and a nonlinear conversion efficiency of 4.7\%~W$^{-1}$, in a LN nanophotonic waveguide only 8~mm long. Our device shows great promise for efficient on-chip wavelength conversion to produce highly-tunable coherent visible light for broad applications, while taking advantage of the mature and cost-effective telecom laser technology.
\end{abstract}

\maketitle 


\section*{Introduction}
Since their invention in the 1960s \cite{Townes58, Maiman60}, lasers have been the backbone of modern optics, playing fundamental roles in optical sciences and technologies. For a coherent light source, wavelength tunability is one of the most important specifications, underlying crucially many applications including optical communications \cite{Agrawal12}, spectroscopy \cite{Hodgkinson12, Vahala16}, frequency metrology \cite{Udem02}, sensing \cite{Vollmer08}, to name some. However, lasing wavebands are naturally limited by gain media. Nonlinear optical parametric processes, such as second-harmonic generation (SHG) and difference-frequency generation (DFG), with the flexible engineering of the phase-matching condition, are probably the most prominent approaches to achieve tunable coherent radiation at optical frequencies that can hardly be obtained by lasers directly \cite{Byer75, Dunn99, Breunig11}.

Lithium niobate (LN), with outstanding nonlinear and linear optical properties, is widely employed for this application, where SFG/DFG has been extensively studied over the past decades, particularly in periodically-poled lithium niobate (PPLN) waveguides \cite{Byer92, Pierce95}. In general, a type-0 configuration is employed to achieve a high conversion efficiency, and temperature tuning is a common technique to vary the operation wavelength of a PPLN waveguide. However, the pump wavelength tunability of the type-0 SHG is fairly limited \cite{Byer92, Pierce95, Fejer97}, mainly due to the relatively small wavelength dependence of the thermo-optic coefficient (although DFG can exhibit a large wavelength tunability with a third wave involved). In fact, LN exhibits a remarkable thermo-optic birefringence \cite{Rendina05, Luo17OL, Luiten18}, significantly greater than most other optical media \cite{Tropf95}. This characteristic can be utilized to greatly increase the wavelength tunability of SHG in PPLN by employing a type-I configuration \cite{Byer92, Pierce95, Cha02}, which, however, inevitably sacrifices seriously the conversion efficiency due to the significantly weaker nonlinearity compared with a type-0 process. 

Over the past decade, a variety of integrated material platforms with $\chi^{(2)}$ nonlinearity have been developed for efficient nonlinear optical parametric processes \cite{Harris06, Vuckovic09, Lipson11, Duchesne11, Tang16Optica, Watts17, Bres17}, where the tight confinement of optical modes is able to significantly enhance nonlinear optical interactions. Recent advances in the integrated LN platform have greatly inspired study of nonlinear optics in LN nanophotonic structures \cite{Hu09, Bowers16, Cheng16, Loncar17OE, Fathpour17APL, Luo17OE, Loncar17NC, Buse17}, showing great potentials for nonlinear wavelength conversion with even higher efficiencies compared with PPLN and other integrated platforms. This provides an opportunity to achieve a large wavelength tunability by using the type-I configuration, while maintaining a high conversion efficiency at the same time.

Here we demonstrate highly-tunable efficient on-chip SHG in a LN nanophotonic waveguide. We achieve SHG through type-I inter-modal phase matching between orthogonal polarizations, and by utilizing the strong thermo-optic birefringence of LN, we demonstrate temperature tuning of the SHG wavelength, with a measured tuning slope of 0.84 nm/K for a telecom pump, almost one order of magnitude higher than that of type-0 SHG in LN \cite{Byer92, Pierce95, Gui10}. Meanwhile, our device is designed to exhibit a large mode overlap, resulting in a theoretical normalized SHG efficiency of $22.2\%~{\rm W^{-1} cm^{-2}}$, which enables us to experimentally demonstrate a conversion efficiency of 4.7$\%~ \rm W^{-1}$ in a waveguide only 8~mm long. Our device is of great promise for efficient on-chip wavelength conversion to produce highly-tunable coherent visible light, which is essential for various integrated photonic applications such as particle and chemical sensing in aqueous environments \cite{Xiao14, Suntivich16, Lu16NC}, while taking advantage of the mature telecom laser technology.

\begin{figure*}[t]
	\centering
	\includegraphics[width=2\columnwidth]{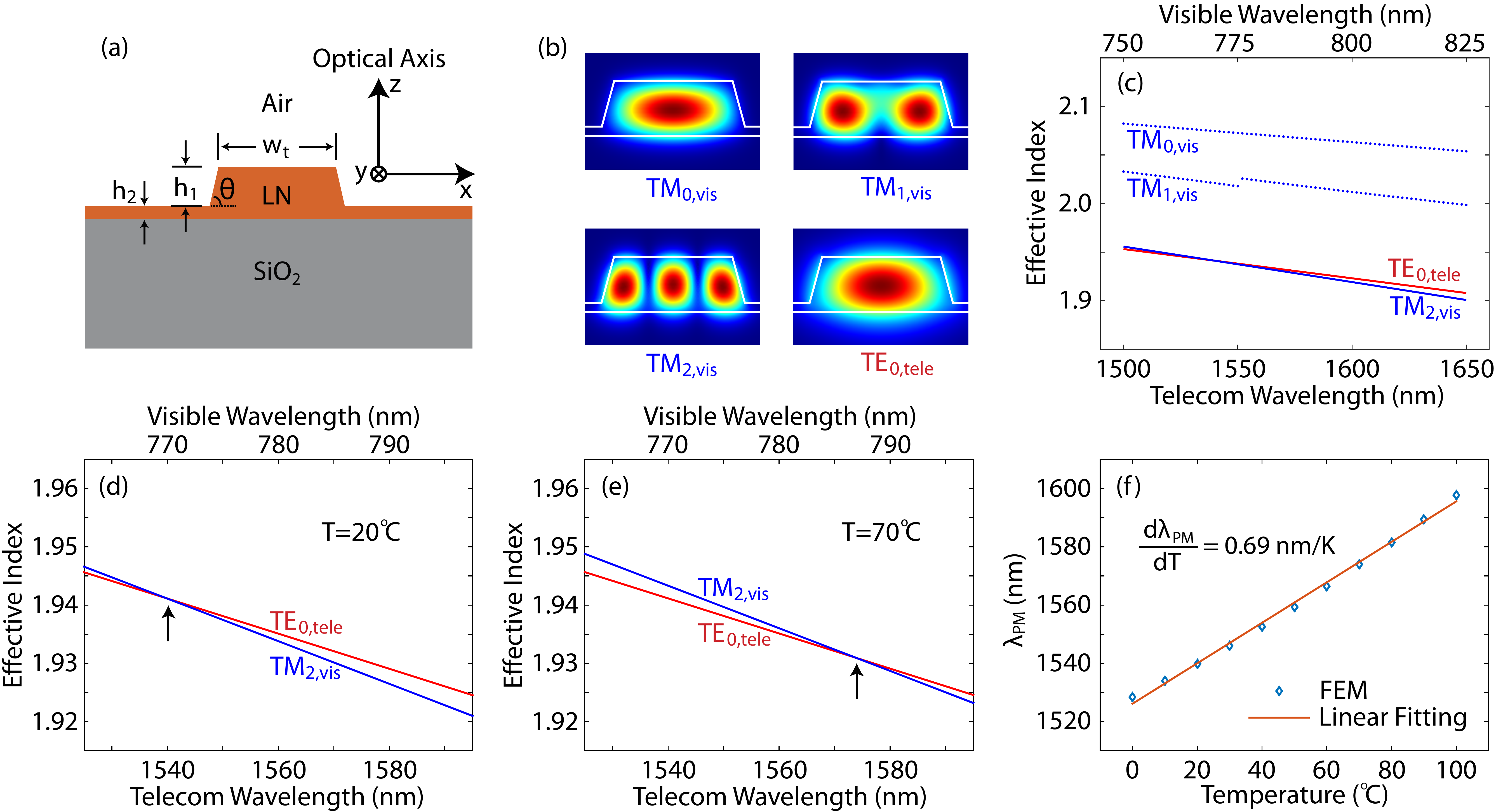}
	\caption{ (a) Schematic of our Z-cut LN waveguide. FEM simulations of (b) mode profiles, and (c) effective indices as functions of wavelength, of TE$_{0,\rm{tele}}$ in the telecom and TM$_{j,\rm{vis}} (j=0,1,2)$ in the visible, where $w_t=1200$ nm, $h_1=460$ nm, $h_2=100$ nm, and $\theta=75^\circ$, at 20$^\circ$C. Discontinuity in the curve of TM$_{1,\rm{vis}}$ is due to its coupling with TE$_{2,\rm{vis}}$ (not shown). Zoom-in of the wavelength-dependent effective indices of TE$_{0,\rm{tele}}$ and TM$_{2,\rm{vis}}$ at (d) 20$^\circ$C, and (e) 70$^\circ$C, with black arrows indicating phase matching. (f) Simulated phase-matched pump wavelength $\lambda_{\rm{PM}}$ as a function of temperature. In (c)-(f), the FEM simulations take into account the temperature and wavelength dependence of the material refractive indices, for both ordinary and extraordinary light \cite{Zelmon97, Rendina05}.}	
	\label{Fig1}
\end{figure*}

\begin{figure*}[t]
	\centering
	\includegraphics[width=1.8\columnwidth]{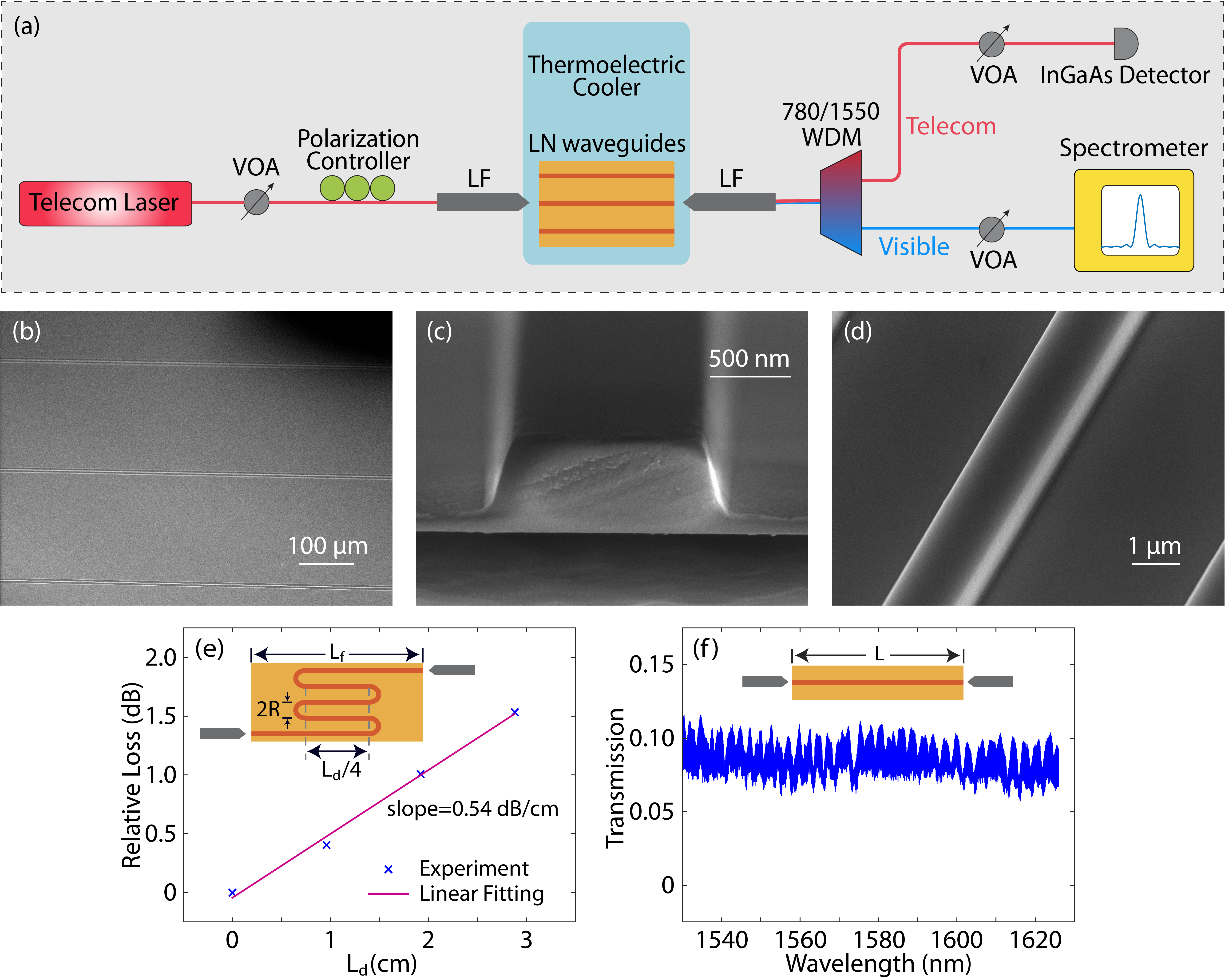}
	\caption{ (a) Experimental setup for device characterization and SHG measurement. Scanning electron microscope pictures showing the waveguide (b) top view, (c) facet, and (d) sidewall. (e) Fiber-to-fiber loss as a function of the differential length $L_d$, relative to that of $L_d=0$, for waveguides schematically illustrated in the inset, where $L_f$ and $R$ are kept as 8 mm and 100 $\mu$m, respectively. (f) Telecom-band transmission spectrum of the TE polarization for a straight waveguide with a length of $L\simeq8$ mm, whose schematic is shown in the inset. VOA: variable optical attenuator; LF: lensed fiber; WDM: wavelength division multiplexer. }
	\label{Fig2}
\end{figure*}

\section*{Waveguide Design}

LN exhibits a significant thermo-optic birefringence, with a value of $|\frac{dn_e}{dT}-\frac{dn_o}{dT}| \sim 4\times10^{-5}\rm{K}^{-1}$ at room temperature \cite{Rendina05, Luo17OL, Luiten18}, where $\frac{dn_e}{dT}$ and $\frac{dn_o}{dT}$ are the thermo-optic coefficients for the extraordinary and ordinary light, respectively. As a result, if SHG occurs in a LN waveguide between optical waves with orthogonal polarizations, a temperature change of the device would result in a considerable variation of the material birefringence which in turn shifts significantly the phase-matched wavelength of the SHG process. In particular, we can maximize this effect by using a Z-cut LN waveguide, which supports ordinarily and extraordinarily polarized optical modes with high polarization purity [see Fig.~\ref{Fig1}(a) and (b)] \cite{Luo17OL}.

We design the geometry of the Z-cut LN waveguide [see Fig.~\ref{Fig1}(a)] such that the fundamental quasi-transverse-electric mode (TE$_{0,\rm{tele}}$) in the telecom band is phase matched with the third-order quasi-transverse-magnetic mode (TM$_{2,\rm{vis}}$) in the visible. Figure \ref{Fig1}(c) and \ref{Fig1}(d) show the effective refractive indices of the two modes, simulated by the finite-element method (FEM), which gives a phase-matched pump wavelength of $\lambda_{\rm{PM}}$=1540~nm at room temperature of $20{\rm ^o C}$. Of particular interest is that LN exhibits a significant thermo-optic effect for extraordinary light ($\frac{dn_{e,vis}}{dT} \approx 4\times10^{-5}\rm{K}^{-1}$) while it is negligible for ordinary light ($\frac{dn_{o,tele}}{dT} \approx 0$) around room temperature \cite{Rendina05}. As a result, when the device temperature increases, the effective refractive index of the TE$_{0,\rm{tele}}$ mode remains nearly intact while that of the TM$_{2,\rm{vis}}$ mode increases considerably. Consequently, the phase-matched wavelength moves dramatically towards longer wavelengths. Figure \ref{Fig1}(e) shows an example, where $\lambda_{\rm{PM}}$ shifts to 1574~nm at a temperature of $70{\rm ^o C}$. Detailed analysis shows that the phase-matched wavelength depends almost linearly on the device temperature, as shown clearly in Fig. \ref{Fig1}(f), with a significant tuning slope of 0.69~nm/K.

Phase matching of the two modes indicates potentially efficient SHG in the designed waveguide. For a lossless waveguide without pump depletion, the SHG efficiency is given by the following expression \cite{BoydBook, Byer75}
\begin{equation}
\Gamma = \frac{P_2}{P_1^2 }=\eta L^2 \left[ \frac{\sin(\Delta L/2)}{\Delta L/2} \right]^2, \label{Gamma}
\end{equation}
where $P_1$ and $P_2$ are the optical powers input at the fundamental wavelength $\lambda$ and produced at the second harmonic, respectively. $L$ is the waveguide length and $\Delta \equiv \frac{4\pi}{\lambda}(n_2-n_1)$ represents the phase mismatch, where $n_1$ and $n_2$ are the effective refractive indices of the TE$_{0,\rm{tele}}$ mode at the fundamental wavelength and the TM$_{2,\rm{vis}}$ mode at the second harmonic, respectively. When the phase-matching condition is satisfied ($\Delta = 0$), Eq.~(\ref{Gamma}) shows the maximum SHG efficiency $\Gamma_0 = \eta L^2$ that depends on the normalized conversion efficiency given as
\begin{equation}
\eta = \frac{8\pi^2}{\epsilon_0 c n_1^2 n_2 \lambda^2} \frac{\zeta^2 d_{\rm eff}^2}{A_{\rm eff}}, \label{eta}
\end{equation}
where $\epsilon_0$ and $c$ are the permittivity and light speed in vacuum, respectively, and $d_{\rm eff}$ is the effective nonlinear susceptibility. In Eq.~(\ref{eta}), $A_{\rm eff} \equiv (A_{1}^2 A_{2})^{\frac{1}{3}}$ is the effective mode area where $A_{i} \equiv \frac{ (\int_{\rm all} |\vec{E}_i|^2 dxdz)^3 }{|\int_{\chi^{(2)}} |\vec{E}_i|^2 \vec{E}_i dxdz|^2}$,~($i=1,2$), and $\zeta$ represents the spatial mode overlap factor between the fundamental and second-harmonic modes, given as
\begin{equation}
\zeta = \frac{ \int_{\chi^{(2)}} (E_{1x}^*)^2 E_{2y} dxdz}{|\int_{\chi^{(2)}}|\vec{E}_1|^2 \vec{E}_1 dxdz|^{\frac{2}{3}}   |\int_{\chi^{(2)}} |\vec{E}_2|^2 \vec{E}_2 dxdz|^{\frac{1}{3}}}, \label{zeta}
\end{equation}
where $\int_{\chi^{(2)}}$ and $\int_{all}$ denote two-dimensional integration over the LN material and all space, respectively, $E_{1x}$ is the x-component of $\vec{E}_1(x,z)$, the electric field of the fundamental mode TE$_{0,\rm{tele}}$, and $E_{2y}$ is the y-component of $\vec{E}_2(x,z)$, the electric field of the second-harmonic mode TM$_{2,\rm{vis}}$.

Equations (\ref{Gamma})-(\ref{zeta}) show that the SHG efficiency depends essentially on the spatial mode overlap $\zeta$, the effective mode area $A_{\rm eff}$, and the effective nonlinear susceptibility $d_{\rm eff}$. Numerical simulation shows that our waveguide exhibits a small $A_{\rm eff} = 1.46~\mu \rm m^2$. In particular, our designed waveguide exhibits a large spatial mode overlap, with $\zeta = 0.32$. As a result, the waveguide exhibits a normalized conversion efficiency as high as $\eta=22.2\%~{\rm W^{-1} cm^{-2}}$. This value is comparable to that of type-0 SHG in typical PPLN \cite{Fejer02, Gui10} and LN nanophotonic waveguides \cite{Loncar17OE} that utilize the maximum component of the $\chi^{(2)}$ nonlinearity ($d_{\rm eff}=d_{33}=27~{\rm pm/V}$), although a type-I configuration is employed here ($d_{\rm eff}=d_{31}=4.3~{\rm pm/V}$ \cite{Roberts92}). In contrast to those type-0 devices, our waveguide is expected to exhibit a significantly larger thermal tuning slope, as we will experimentally demonstrate in the following.


\section*{Experimental Results}

\begin{figure*}[t]
	\centering
	\includegraphics[width=2\columnwidth]{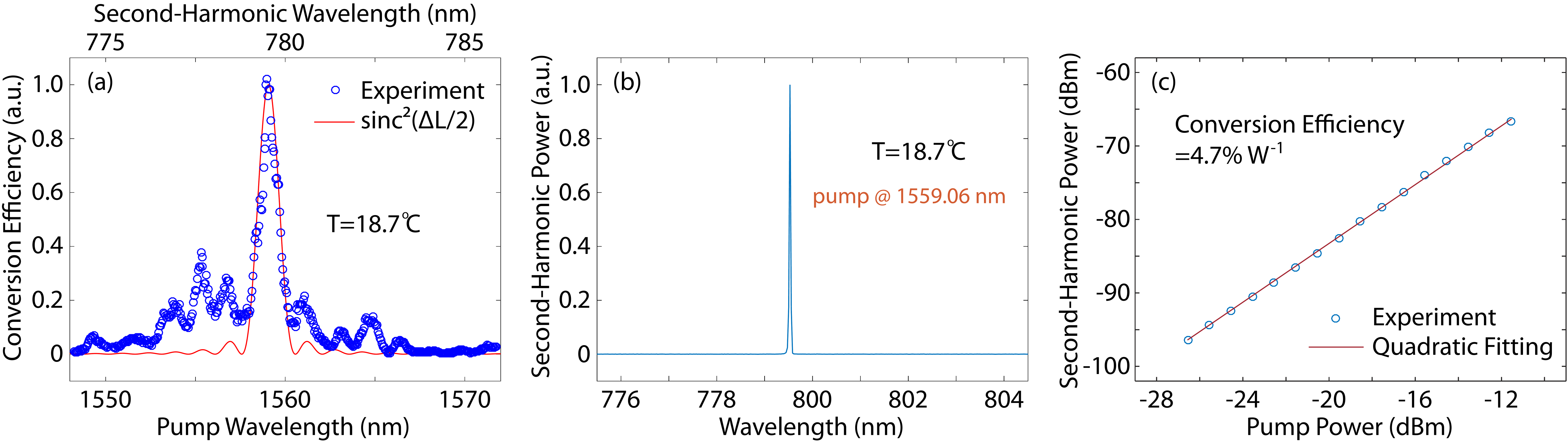}
	\caption{ SHG from a straight LN nanophotonic waveguide with a length of 8 mm. (a) Conversion efficiency spectrum at T=18.7$^\circ$C, with the center wavelength of the sinc$^2$-function aligned to the measured peak. (b) SHG spectrum at a fixed pump wavelength of 1559.06 nm at T=18.7$^\circ$C. (c) Second-harmonic power as a function of pump power, with experimental data compared with a quadratic fitting, exhibiting a conversion efficiency of 4.7$\%~{\rm W^{-1}}$. }
	\label{Fig3}
\end{figure*}

\begin{figure*}[t]
	\centering
	\includegraphics[width=1.9\columnwidth]{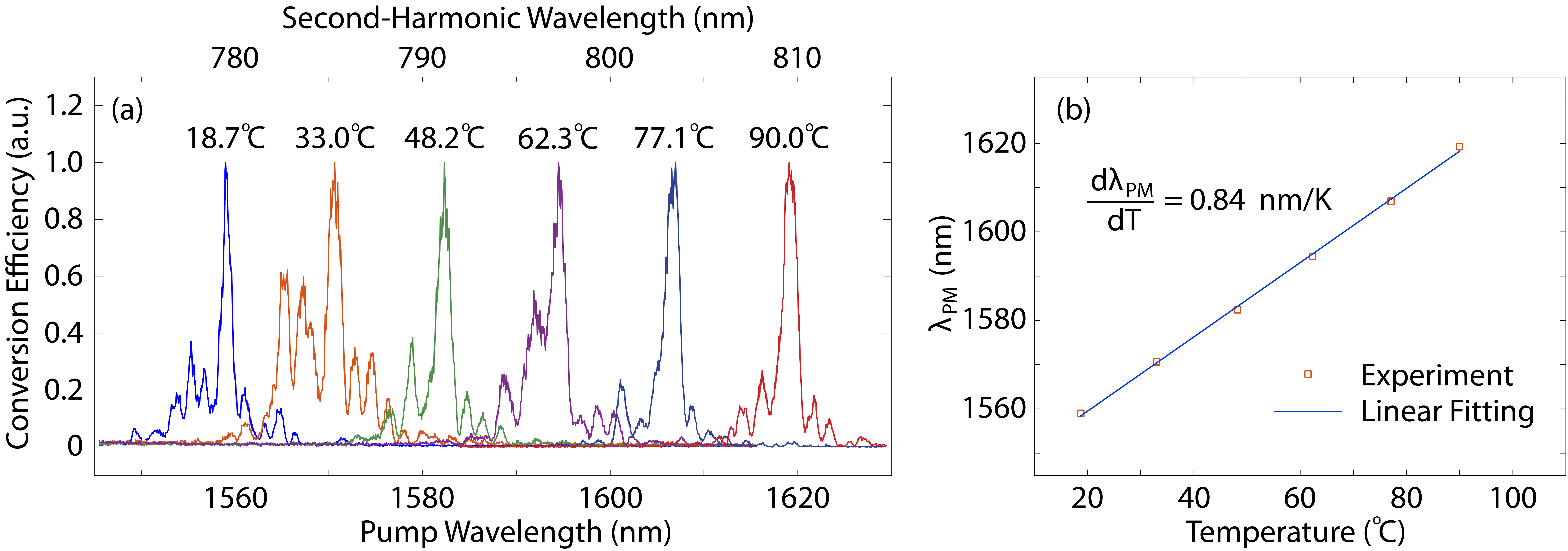}
	\caption{ Thermal tuning of SHG. (a) Conversion efficiency spectra at different temperatures. (b) Measured phase-matched pump wavelength $\lambda_{\rm{PM}}$ as a function of temperature. }
	\label{Fig4}
\end{figure*}

To confirm our simulation results, we fabricated waveguides on a Z-cut LN-on-insulator wafer [see Fig.~\ref{Fig2}(b)], where the LN thin film has a thickness of $\sim$560 nm, sitting on 2-$\mu$m-thick buried oxide. Figure \ref{Fig2}(c) shows the cross-section of a fabricated waveguide, whose geometry is very close to our design [see Fig.\ref{Fig1}(a)]. In particular, as presented in Fig.~\ref{Fig2}(d), the waveguide sidewall is very smooth, implying a low propagation loss. In order to quantify the propagation and coupling losses, we fabricated waveguides with the same cross-section but different lengths, as schematically shown in the inset of Fig.~\ref{Fig2}(e). Since these waveguides share the same coupling and bending losses, by measuring their transmission as a function of the differential length, we can extract the propagation loss. Figure \ref{Fig2}(e) shows the measurement results, where the propagation loss of straight waveguides for the TE$_{0,\rm{tele}}$ mode is measured to be 0.54 dB/cm, a small value that represents the state-of-the-art quality of LN nanophotonic waveguides. Together with the overall fiber-to-fiber transmission of a straight waveguide [for example, see Fig.~\ref{Fig2}(f)], we obtained a fiber-to-chip coupling loss of about 5 dB/facet.

To demonstrate SHG, we employed a straight waveguide with a length of about 8~mm. We launched a telecom-band continuous-wave (CW) laser into the device, with the setup shown in Fig.~\ref{Fig2}(a). By scanning the laser, we were able to measure the efficiency spectrum of SHG. One example is shown in Fig.~\ref{Fig3}(a), which shows a phase-matched pump wavelength of 1559~nm at a temperature of 18.7$^\circ$C. The main lobe of the recorded efficiency spectrum agrees well with the theoretical expectation from the function ${\rm sinc}^2(\Delta L/2)=[\frac{\sin(\Delta L/2)}{\Delta L/2}]^2$. The strong side lobes are likely caused by reflection at facets that are not perfectly smooth.

By fixing the pump wavelength at 1559.06 nm, which exhibits the peak conversion efficiency, we observed coherent radiation from its second harmonic at 779.53 nm [see Fig.~\ref{Fig3}(b)]. The second harmonic shows a quadratic power dependence on the pump, which agrees very well with the theoretical expectation. Fitting the experimental data, we obtained an on-chip conversion efficiency of 4.7\%~$\rm W^{-1}$ [see Fig.~\ref{Fig3}(c)]. This value is smaller than the theoretical upper limit given by $\eta L^2$ (=14.2\%~$\rm W^{-1}$), mainly due to the non-zero propagation losses at both wavelengths.

To show the spectral tuning capability of our device, we varied the device temperature from 18.7${\rm ^o C}$ to 90.0${\rm ^o C}$ and measured the SHG efficiency spectra. Figure \ref{Fig4}(a) presents the recorded spectra at different temperatures. It shows clearly that the SHG spectrum shifts towards longer wavelengths when the device temperature increases. The spectral shape also changes with temperature, potentially due to the temperature-dependent facet reflection. By mapping the phase-matched pump wavelength as a function of temperature, we obtained Fig.~\ref{Fig4}(b), showing an experimentally measured tuning slope of $\frac{d\lambda_{\rm{PM}}}{dT}=0.84$ nm/K, almost one order of magnitude larger than that achieved by type-0 SHG in LN \cite{Byer92, Pierce95,Gui10}. The experimental results agree well with our simulations [see Fig.~\ref{Fig1}(f)]. A slightly larger experimental value of the tuning slope is likely contributed by pyroelectric \cite{Bernal13} and thermal expansion effects \cite{Smith69} in the waveguide cross-section, which were not taken into account in the simulations.


\section*{Conclusion}
In conclusion, we have demonstrated highly-tunable efficient SHG in a LN nanophotonic waveguide. The LN waveguide exhibits a high optical quality with a propagation loss as low as 0.54 dB/cm in the telecom band, which represents the state-of-the-art quality of LN nanophotonic waveguides reported to date \cite{Loncar17Optica, Peruzzo18, Buse17}. In particular, we took advantage of the strong thermo-optic birefringence of LN to achieve thermal tuning of the SHG wavelength, with a tuning slope of 0.84 nm/K for a telecom-band pump, significantly higher than that offered by type-0 SHG in LN. At the same time, thanks to the tight mode confinement and a large spatial mode overlap, our waveguide exhibits a high theoretical normalized conversion efficiency of $22.2\%~{\rm W^{-1} cm^{-2}}$ even for the type-I inter-modal phase matching, which is comparable to that of type-0 SHG in typical PPLN and LN nanophotonic waveguides utilizing the largest nonlinear term $d_{33}$. Our waveguide design enabled us to experimentally record a SHG efficiency of 4.7$\%~\rm W^{-1}$ inside a waveguide only 8~mm long. Our device shows great promise for on-chip wavelength conversion that takes advantage of the mature telecom-band laser technology to produce highly-tunable coherent light in the visible.

\section*{Appendices}

\subsection{Device Fabrication} Starting from a Z-cut LN-on-insulator wafer by NANOLN, we used electron-beam lithography with ZEP520A as the resist for device patterning, and Argon ion milling for etching. Next, in order to remove the remaining resist and material residuals, we treated the chip with oxygen plasma followed by diluted hydrofluoric acid. Finally, we diced the chip and polished the facets for light coupling.

\subsection{Experimental Setup} Pump light from a CW tunable telecom-band laser was coupled via a lensed fiber into the device chip, which was placed on top of a thermoelectric cooler that controls the temperature. At the waveguide output, pump light was collected together with the frequency-doubled light by a second lensed fiber. After being separated from its second harmonic by a 780/1550 wavelength division multiplexer, the telecom pump light was directed to an InGaAs detector for characterization, while the generated visible light was sent to a spectrometer for detection. A fiber polarization controller was used for optimal coupling from the input lensed fiber to the wanted waveguide mode, and variable optical attenuators were employed to study the power dependence of SHG. The spectrometer was cooled by liquid nitrogen for a high sensitivity.

\subsection{SHG Spectrum Measurement} After aligning lensed fibers to the waveguide for optimal coupling, we scanned the telecom-band pump laser, with the spectrometer recording generated second-harmonic light during the whole laser scanning period. This process was repeated for different temperatures, which were controlled by the thermoelectric cooler under the device chip, to obtain the temperature dependence of the SHG spectrum.

\section*{Funding}
This work was supported in part by the National Science Foundation under Grant No.~ECCS-1641099 and ECCS-1509749, and by the Defense Advanced Research Projects Agency SCOUT program through Grant No.~W31P4Q-15-1-0007 from the U.S. Army Aviation and Missile Research, Development, and Engineering Center (AMRDEC). The views and conclusions contained in this document are those of the authors and should not be interpreted as representing the official policies, either expressed or implied, of the Defense Advanced Research Projects Agency, the U.S. Army, or the U.S. Government.

\section*{Acknowledgment}
The authors thank Chengyu Liu at Cornell University for helpful discussions on fabrication. This work was performed in part at the Cornell NanoScale Facility, a member of the National Nanotechnology Coordinated Infrastructure (National Science Foundation, ECCS-1542081), and at the Cornell Center for Materials Research (National Science Foundation, DMR-1719875).

\bibliography{References}

\end{document}